\documentclass[10pt,conference]{IEEEtran}
\IEEEoverridecommandlockouts
\usepackage{cite}
\usepackage{amsmath,amssymb,amsfonts}
\usepackage{algorithmic}
\usepackage{graphicx}
\usepackage{textcomp}
\usepackage{xcolor}
\usepackage{graphicx}
\usepackage{float}
\usepackage{algorithmic}
\usepackage{subfigure,multirow,bm,stfloats}

\def\BibTeX{{\rm B\kern-.05em{\sc i\kern-.025em b}\kern-.08em
    T\kern-.1667em\lower.7ex\hbox{E}\kern-.125emX}}
    
\begin{document}

\title{Exemplar-Based Radio Map Reconstruction of Missing Areas Using Propagation Priority
}

\author{
\IEEEauthorblockA{Songyang Zhang*, Tianhang Yu*,
Jonathan Tivald*, Brian Choi$^\dagger$, Feng Ouyang$^\dagger$, Zhi Ding*, \textit{Fellow, IEEE}\\
*Department of Electrical and Computer Engineering, 
{University of California}, Davis, CA, USA, 95616 \\
$^\dagger$Applied Physics Laboratory,
Johns Hopkins University,
Laurel, Maryland, USA, 20723\\
E-mail: \{sydzhang, thgyu, jrtivald, zding\}@ucdavis.edu, \{Brian.Choi, Feng.Ouyang\}@jhuapl.edu}
\IEEEcompsocitemizethanks{
		\IEEEcompsocthanksitem 
		 This work was supported in part by the National Science Foundation under Grant 2029848 and Grant 2029027; and in part by Johns Hopkins University Applied Physics Laboratory Independent Research and Development Funds.
	}    
}
\maketitle

\begin{abstract}
Radio map describes network coverage and is a practically
important tool for network planning in modern wireless systems. Generally, radio strength
measurements are collected
to construct fine-resolution radio maps for analysis. However, 
certain protected 
areas are not accessible
for measurement due to 
physical constraints and
security considerations, 
leading to blanked spaces on
a radio map. 
Non-uniformly spaced measurement
and uneven observation resolution
make it more difficult for radio map estimation and spectrum planning in protected areas. This work explores the distribution of radio spectrum strengths and proposes an exemplar-based approach to reconstruct 
missing areas on a radio map. 
Instead of taking generic image processing approaches, we
leverage radio propagation models to determine directions of region filling and develop two different schemes to estimate the missing
radio signal power. Our test results based on high-fidelity simulation demonstrate 
efficacy of the proposed methods for radio map reconstruction. 
\end{abstract}


\section{Introduction}
With increasingly expansion of sensor network and Internet of Things (IoT) deployment, allocation of radio spectrum is becoming more complex and dynamic, and poses further challenges in managing radio resources and enabling new applications \cite{b1}. To better capture  spectrum usage pattern and improve efficiency of resource management, radio maps can play more important roles in the modern wireless communication systems. A radio map is generally characterized by the power spectral density (PSD) over geographical locations, frequencies and time \cite{b2}. Providing rich and useful information regarding spectrum activities and propagation channels, radio maps 
can provide information on detailed PSD distribution and help develop spectrum management applications \cite{b3}. Usually, a high-resolution radio map should be constructed from sparser measurements \cite{b4}. 
One major challenge lies in reconstructing more complete radio maps from partial observations.

General construction of radio maps utilizes either model-based methods or model-free methods \cite{b2}. Model-based methods
assume certain signal propagation models 
to express the received PSD as a combination of transmitted PSD from active transmitters. For example, an interpolation method 
\cite{b5} proposes to utilize log-distance path loss model (LDPL) for Wi-fi radio map reconstruction. In \cite{b4}, another model-based method introduces 
the use of thin-plate splines kernels. Different from model-based approaches, model-free methods do not rely on specific signal propagation models but favor neighborhood information. Typical examples include inverse distance weighted (IDW) interpolation \cite{b6}, Kriging-based interpolation \cite{b7} and Radial Basis
Functions (RBF) interpolation \cite{b8}. In addition, graph-based approaches, such as graph signal processing \cite{b9} and label propagation \cite{b10}, can also assist
radio map reconstruction. Beyond interpolation-based methods, machine learning has also attracted significant attention in radio map reconstruction owing to its 
ability to utilize hidden data features \cite{b11,b12,b13}.

\begin{figure}[t]
	\centering
	\subfigure[]{
		\label{saml1}
		%\hspace{1in}
		\includegraphics[height=2.2cm]{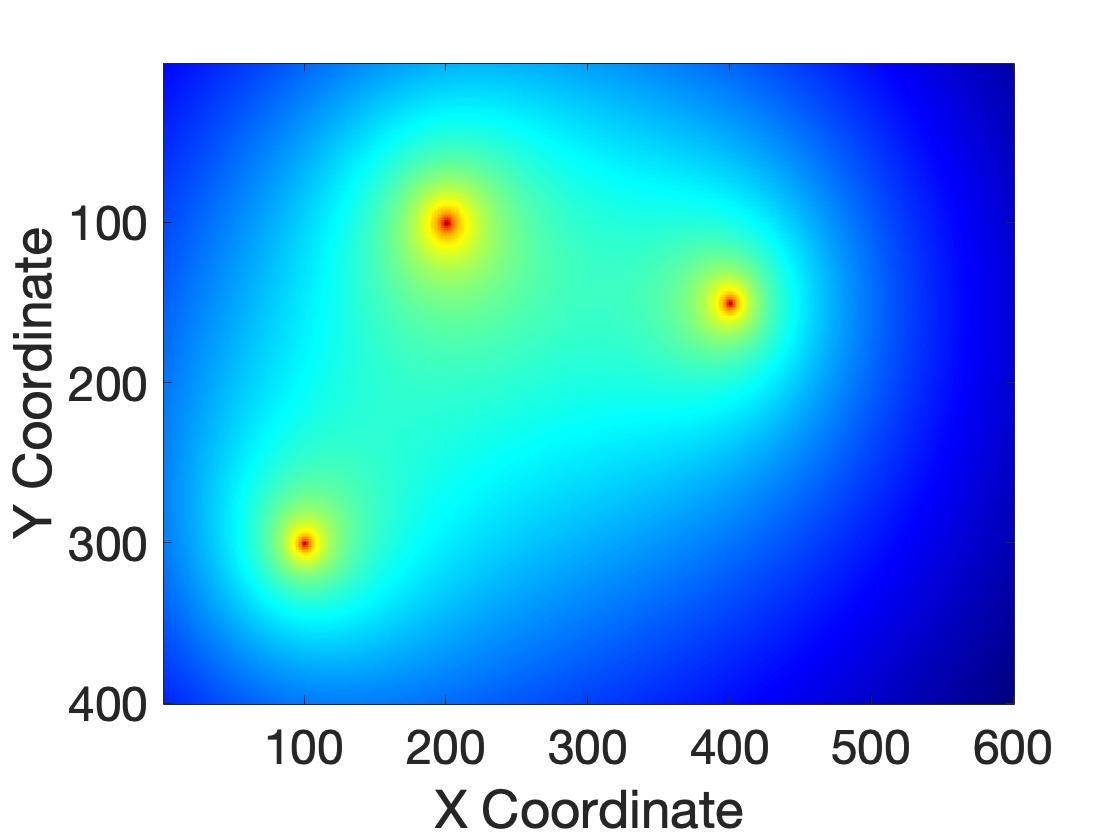}}
	\hspace{1cm}
	\subfigure[]{
		\label{same1}
		\includegraphics[height=2.2cm]{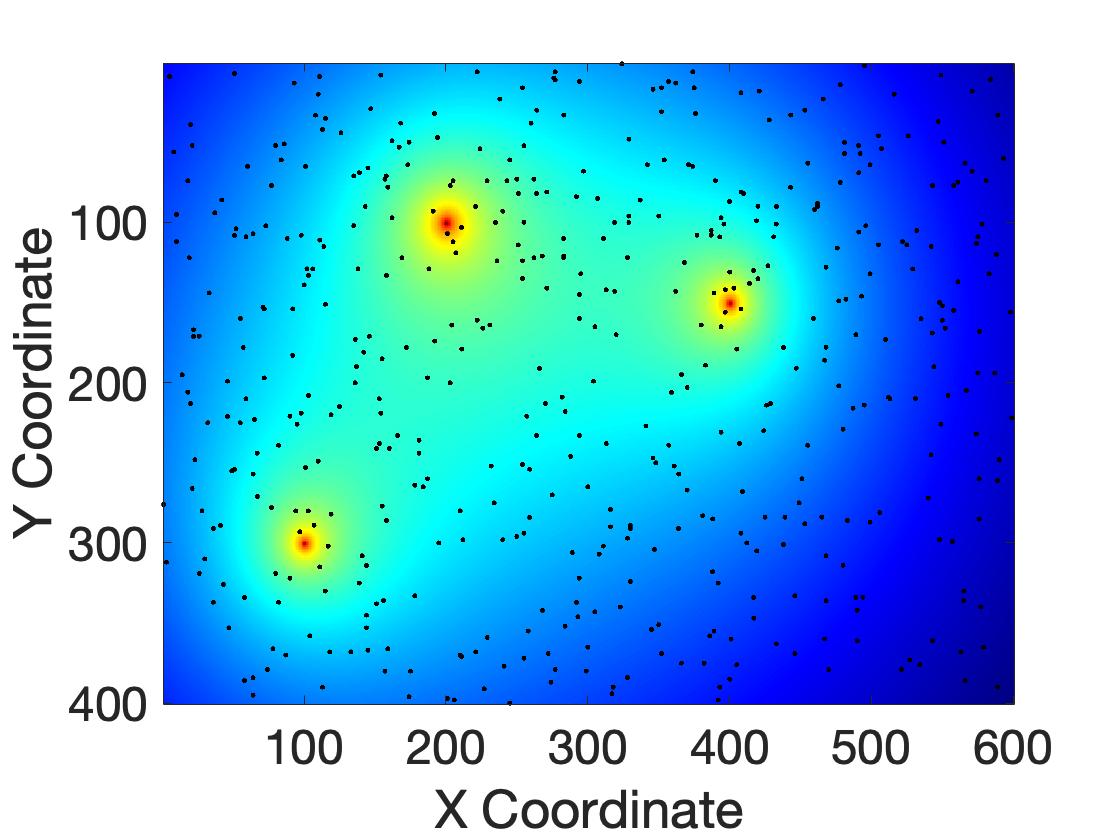}}\\
	\vspace{-3mm}
	\subfigure[]{
		\label{samb11}
		\includegraphics[height=2.2cm]{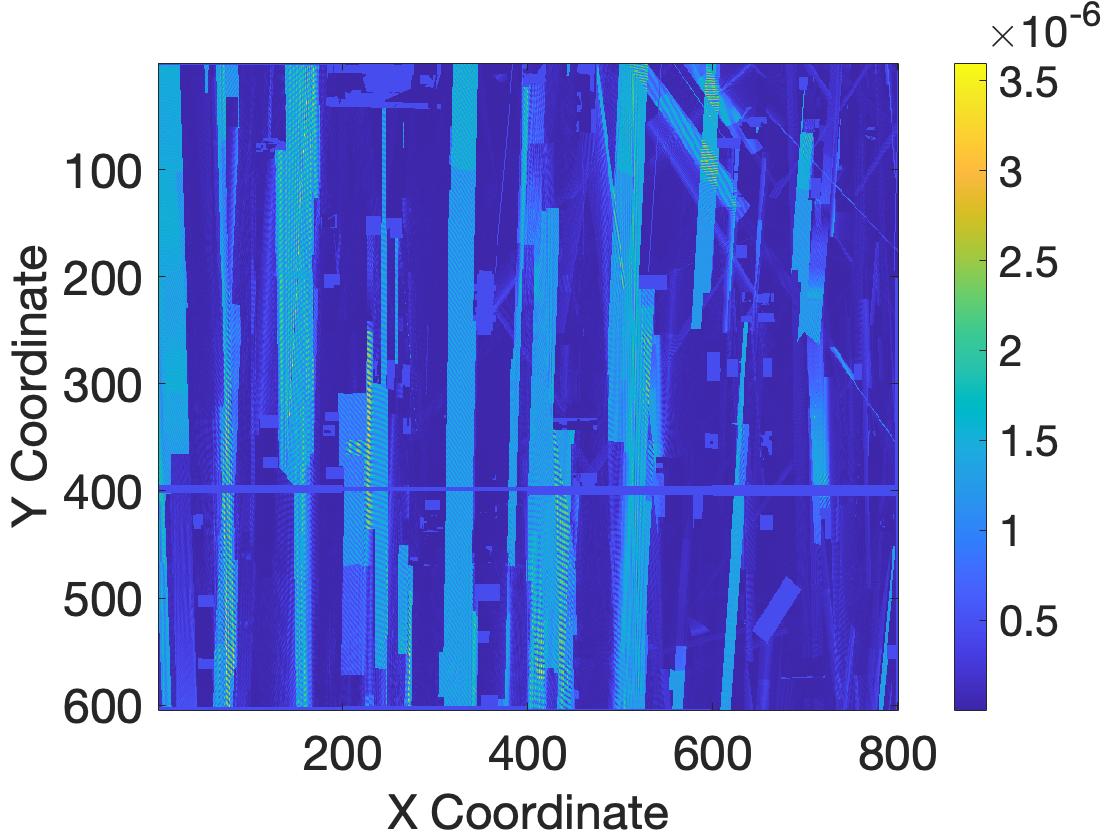}}
	\hspace{1cm}
	\subfigure[]{
		\label{samb21}
		\includegraphics[height=2.2cm,width=2.7cm]{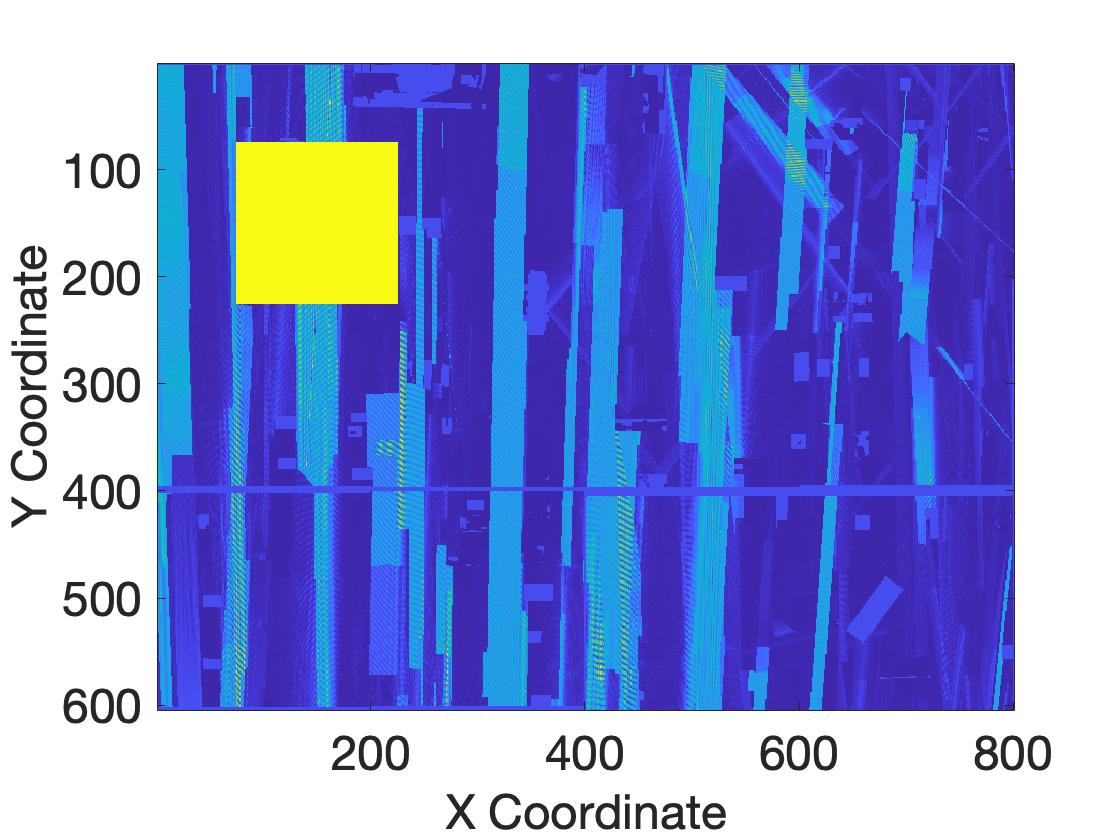}}
	\caption{Examples of Radio Map: Figure (a) and (b) show the power spectrum density and sensor (receivers represented by dots) placement for general large-scale radio map; Figure (c) and (d) show the spectrum distribution (Watts) and missing observations for restricted areas (marked in yellow) of a small-scale radio map (e.g. several street blocks), which only covers a small part of the large-scale radio map. Note that, the coordinates here are the index of PSD conformed to the grid. Usually, the small-scale radio map has higher resolution and smaller area than the large-scale ones.
	}
	\vspace{-6mm}
	\label{samex1}
\end{figure}

Presently, most existing approaches focus on  constructing radio maps from sparse observations, where sensors are spread
over a given region as shown in 
Fig.~\ref{same1}. However, in 
cases involving inaccessible,
restricted, or protected areas, radio
measurement is not available,
leading to missing observations 
of certain regions or blocks. The radio map construction for such restricted areas is 
more challenging and does not lend themselves to traditional radiomap 
construction methods. 
First, unlike large-scale radio map, missing
observations of power spectrum covering
restricted areas occur in relatively smaller regions, such as the example of Fig. \ref{samb21}.  PSD distribution in these small-scale regions tends not
to follow well known propagation models but
is more sensitive to small scale environmental
features, which makes the implementation of model-based methods more difficult. 
Secondly, since available measurement
samples are uneven and observations of 
some entire segments are missing,
interpolation methods are ineffective
without accurate and reliable
neighborhood information, 
especially for restricted regions.
Last but not least, for practical
reasons, observed data are usually limited, providing insufficient training samples for learning-based approaches.

In this work, to capture the spectrum power distribution from limited and uneven observations, we propose an exemplar-based approach using radio propagation priority to reconstruct radio map in restricted
or inaccessible areas. The main contributions of our work are summarized as follows:
\begin{itemize}
	\item Through exploring the pattern of spectrum power from observed data and integrating radio propagation models, we introduce  propagation model-based priority to define   directions of data filling for missing regions.
	\item By analyzing correlations from observed signals, we propose to estimate 
	missing radio PSD values based on exemplar copying and dictionary learning, respectively.
\end{itemize}
We compare our proposed methods with traditional radio map constructions
by testing over a Applied Physics Laboratory (APL) dataset
from Johns Hopkins University (JHU). Our
test results demonstrate the effectiveness 
of the proposed radio map reconstruction method for restricted/inaccessible areas.

\begin{figure}[t]
	\centering
	\includegraphics[width=2.8in]{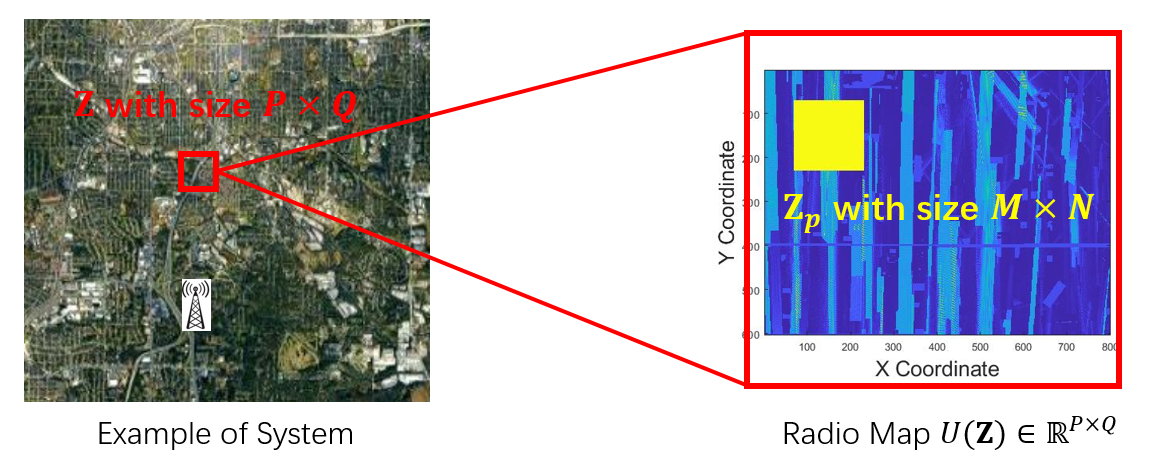}
	\vspace{-2mm}
	\caption{Illustration of Objective Scenarios: the restricted/inaccessible area $\mathbf{Z}_p$ is marked in yellow with area size $M\times N$ in a small-scale radio map $U(\mathbf{Z})\in\mathbb{R}^{P \times Q}$.}
	\label{exm}
	\vspace{-6mm}
\end{figure}

\section{Problem Description}\label{sysm}
Our model considers a wireless network coverage of
a rectangular area with one transmitter. 
All radio observations are arranged on a regular grid and are located in the rectangular area $\mathbf{Z}$ with size $P\times Q$ within the network coverage, denoted by ${U}(\mathbf{Z})\in\mathbb{R}^{P\times Q}$ shown as Fig. \ref{exm}.
Here, $P$ and $Q$ are the size of grid.
Each observation in ${U}(\mathbf{Z})$ is characterized by a 2-dimensional (2D) coordinates $Z_i=(X_i,Y_i)$ and the corresponding radio spectrum power $e_i=U(Z_i)$. The restricted/inaccessible area $\mathbf{Z}_p$ with size $M\times N$ is located within $\mathbf{Z}$, marked
as yellow in Fig. \ref{exm}, where $M\leq P, N\leq Q$. No observation within $\mathbf{Z}_p$ is available. Compared to traditional 
radio map reconstruction problems, the small-scale radio map has higher resolution (e.g., accurate to 1 meter) and smaller area, which make it more sensitive to the nearby environment, such as buildings, trees and roads. Moreover, we only have limited and unbalanced observations around restricted/inaccessible areas.
Our goal is to estimate $U(\mathbf{Z}_p)$ in the restricted/inaccessible area $\mathbf{Z}_p$ from other observed samples in area $\mathbf{Z}$.

Although we only consider one transmitter here, our framework can be directly extended to multiple transmitters by combining all the transmitted PSD. For convenience, we will focus on the one transmitter case in this work and leave more detailed analysis in future works. 

Note that, our objective here is similar to the image inpainting \cite{b14} problem in computer vision. However, the traditional image inpainting only concerns about pixel values 
but not the wireless communication context. Hence, it is ineffective for capturing spectrum power distribution in radio scenarios. 
Besides observed values, we also consider the radio propagation model to assist radio map reconstruction for restricted/inaccessible areas. More analysis and comparison will be discussed in Section \ref{exp}.

\section{Exemplar-Based Radio Map Reconstruction}
In this section, we introduce an exemplar-based radio map reconstruction using radio propagation priority.

\begin{figure}[t]
	\centering
	\includegraphics[width=3.5in]{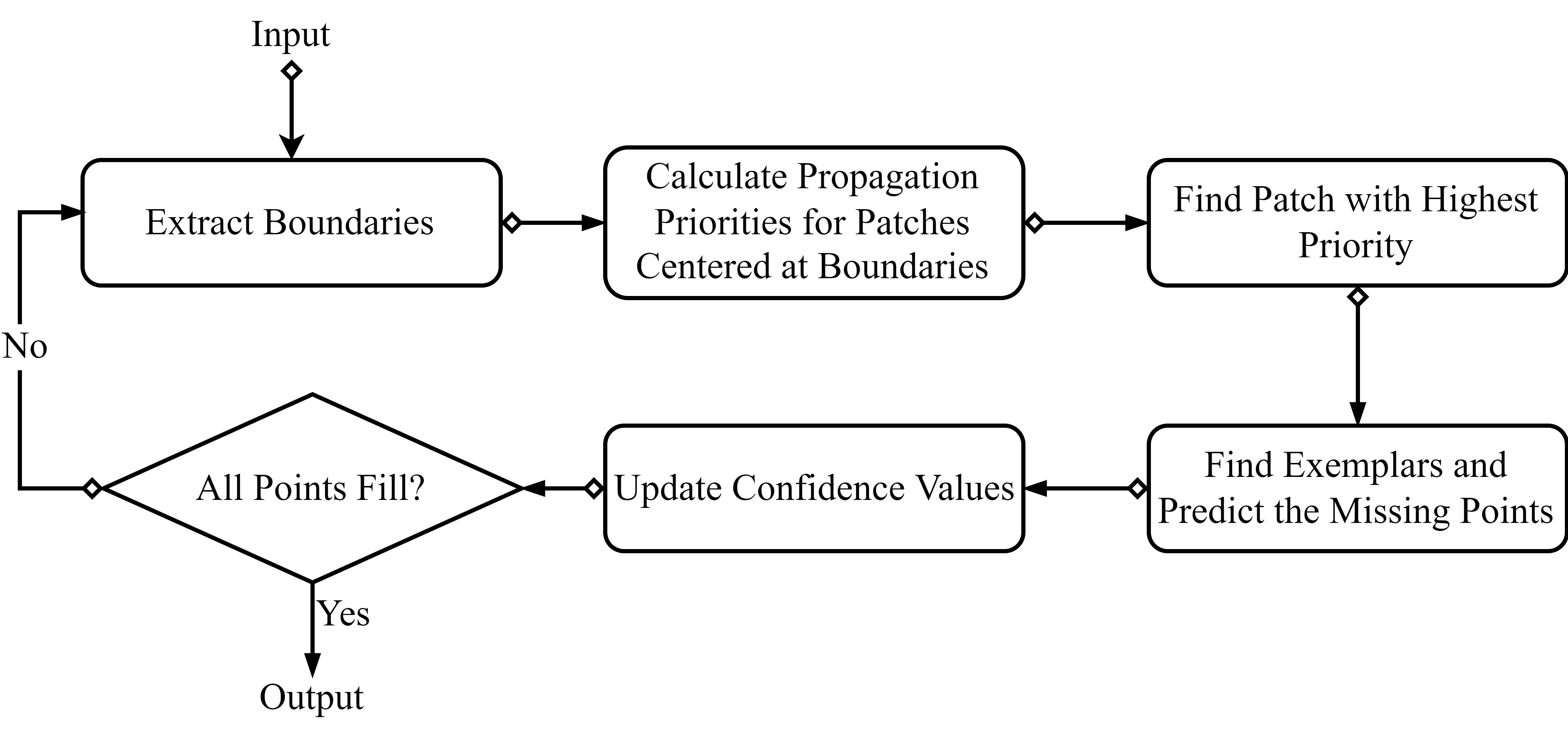}
	\vspace{-8mm}
	\caption{Scheme of Proposed Method}
	\label{sch}
	\vspace{-3mm}
\end{figure}

\begin{figure}[t]
	\centering
	\subfigure[]{
		\label{exem1}
		\includegraphics[height=2.1cm]{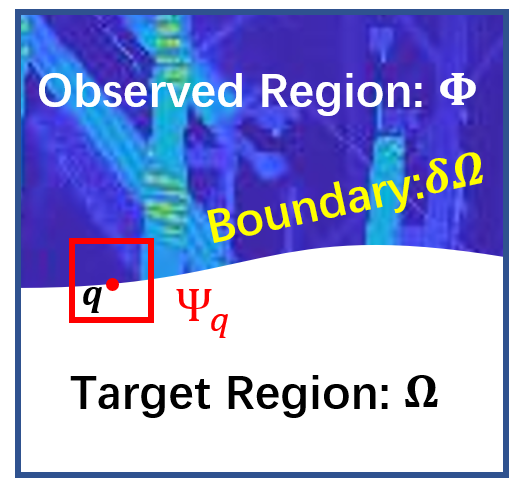}}
	\hspace{1.5cm}
	\subfigure[]{
		\label{exem2}
		\includegraphics[height=2.1cm]{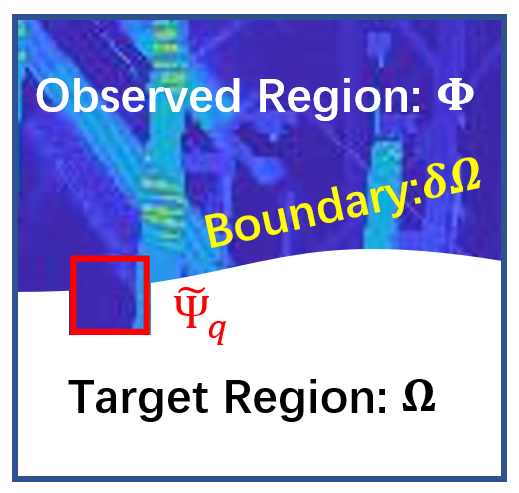}}
	\vspace{-2mm}
	\caption{Illustration of Filling Process: a) Select a patch $\Psi_q$ in the boundary $\delta \Omega$; b) Estimate the missing values in $\Psi_q$ and regenerate $\tilde\Psi_q$}
	\label{exeme}
	\vspace{-5mm}
\end{figure}

\subsection{Overview of the Proposed Method}
To fill a region based on 
surrounding observations, one intuitive way is to estimate the missing values patch (block) by patch (block) from boundaries between observed and target (restricted area) regions to the center of the restricted/inaccessible area.
In this work, we follow a similar scheme to reconstruct the radio map from observations as shown in Fig. \ref{sch}. To estimate radio power in restricted areas, we start from a small selected patch centered at the boundaries shown as Fig. \ref{exeme}. Next, we estimate the missing values for this selected patch and update the boundary. Through patch-by-patch estimation of the missing values, we can obtain the reconstructed radio map for the whole restricted/inaccessible area. The general steps are described as follows.
\begin{itemize}
	\item Step 1: Extract the boundary $\delta \Omega$ between observed region $\Phi$ and target region $\Omega$ (initialized as $\mathbf{Z}_p$) in $\mathbf{Z}$;
	\item Step 2: Given a patch $\Psi_p$ with size $n\times n$ centered at point $p$ located at boundaries, i.e., $p\in\delta \Omega$, calculate the priority of the patch as $P(p)$ based on the texture properties of observations and radio propagation features;
	\item Step 3: Order all patches $\Psi_p$ centered at $\delta \Omega$ by $P(p)$ and select the one with highest priority as $\Psi_q$;
	\item Step 4: Select exemplars from observed region for $\Psi_q$ and estimate the missing values in $\Psi_q$;
	\item Step 5: Update $\Phi$ and $\Omega$;
	\item Step 6: Update the confidence term in the priority;
	\item Step 7: Repeat Step 1-6 until all the missing values in the restricted/inaccessible area $\mathbf{Z}_p$ are estimated.
\end{itemize}
From the steps above, the key issues in the proposed method are how to define priority $P(p)$ to determine the filling direction, and how to estimate the missing values from exemplars. We will discuss more details in Section \ref{detailpro}.

\subsection{Details in the Proposed Method} \label{detailpro}
This part introduces definition of priority based on radio propagation and two approaches to estimate the missing radio map values.

\subsubsection{Definition of Priority}
To find a suitable direction of filling the missing region, we expect to propagate the key information in texture and radio spectrum with larger certainty. Thus, we define the priority of patch selection as follows:

\begin{figure}[t]
	\centering
	\subfigure[Data term.]{
		\label{pp1}
		\includegraphics[height=2.5cm]{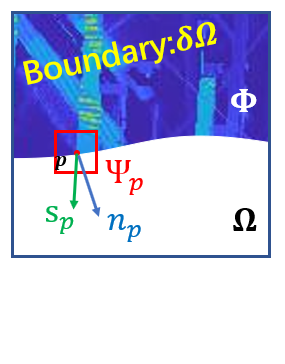}}
	\hspace{3mm}
	\subfigure[Radio term.]{
		\label{pp2}
		\includegraphics[height=2.5cm]{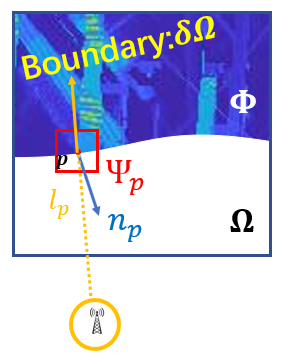}}
	\hspace{1mm}
	\subfigure[Block term.]{
		\label{pp3}
		\includegraphics[height=2.5cm]{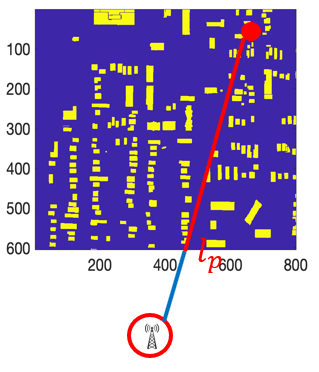}}
	\vspace{-2mm}
	\caption{Illustration of Calculating Priority}
	\label{pp}
	\vspace{-5mm}
\end{figure}

\begin{equation}
	P(p)=C(p)\cdot D(p)\cdot B(p)\cdot L(p),
\end{equation}
where the confidence term $C(p)$ together with data term $D(p)$ contain radio map pattern information (texture), whereas radio propagation term $L(p)$ together with block term $B(p)$ describe radio propagation properties. More specifically:
\begin{itemize}
	\item $C(p)$: The confidence term $C(p)$ describes the confidence level of the PSD within $\Psi_p$. If there are more points from the observed region, the corresponding patch has a higher confidence value. Suppose that there are $n\times n$ points in $\Psi_p$. The confidence term is calculated as
	\begin{equation}
		C(p)=\frac{\sum_{v \in (\Psi_p\cap \Phi)}C(v)}{n\times n},
	\end{equation}
	where $C(v)$ is initialized as $C(v)=1$ for $v\in \Phi$; otherwise, $C(v)=0$. For each iteration, confidence term $C(u)$ for a newly-filled point $u$ in $\tilde \Psi_q$ is updated by $C(u)=C(q)$ before the next iteration at Step 6.
	
	\item $D(p)$: $D(p)$ is the data term describing the gradients of texture. Suppose that the normal of boundary at $p$ is $\mathbf{n}_p$, and the orthogonal direction of the texture gradient at $p$ is $\mathbf{s}_p=\nabla T_p^{\perp}$ where $T_p$ is the power level around $p$, and $^{\perp}$ is the orthogonal operator. The data term is defined as 
	\begin{equation}
		D(p)=\frac{|\mathbf{s}_p \cdot\mathbf{n}_p|}{\alpha},
	\end{equation}
	where $\cdot$ is the inner product, and $\alpha$ is a normalization factor (e.g., $\alpha=1$ if $\mathbf{n}_p$ and $\mathbf{s}_p$ are unit vectors). The data term describes the intensity of radio map texture hitting the boundaries.
	
	\item $L(p)$: The radio propagation term describes the relationship between the PSD at $p$ and the transmitter at location $t$. In model-based approaches, signal power is  a function of distance to the transmitter \cite{b5,b6}. Similarly, we embed the power strength information in $L(p)$ based on the distance $d(t,p)$ between $t$ and $p$. Since  radio propagation property is similar to the texture change described in data term $D(p)$, we can also measure the certainty of radio propagation based on its strengths hitting the boundary:
	\begin{equation}
		L(p)=|d(t,p)|^{-\beta}|\mathbf{l}_p\cdot \mathbf{n}_p|,
	\end{equation}
	where $\beta$ is the inverse distance parameter, $\mathbf{n}_p$ is the normal of boundary at $p$ and $\mathbf{l}_p$ is the direction of radio propagation from $t$ to $p$, shown as Fig. \ref{pp2}.
	
	\item $B(p)$: Since radio map around a restricted/inaccessible area is small-scale and sensitive to the environment, we could also embed information of propagation obstacles in block term $B(p)$. From additional resources, such as satellite image and city map, we can segment buildings (in yellow) and background (in blue) as shown in Fig. \ref{pp3}. Let $l_p$ be part of the line connecting $t$ and $p$ within the whole region $\mathbf{Z}$ defined in Section \ref{sysm}, i.e., red parts in Fig. \ref{pp3}. Then we define $B(p)$ as 
	\begin{equation}\label{btm}
		B(p)=1-\frac{\textnormal{the length covering buildings in } l_p}{\textnormal{the total length of } l_p}.
	\end{equation}
	If the radio propagates over more obstacles, $B_p$ is smaller and the priority would be reduced.
\end{itemize}
By selecting those patches with the
largest $P(p)$ to fill first, we can determine the filling direction with larger confidence level in both texture and radio propagation. Note that we provide the priority based on single transmitter here. If there are multiple transmitters, one can simply modify $P(p)$ as
\begin{equation}
	P(p)=C(p)\cdot D(p)\cdot [\sum_i B_i(p)\cdot L_i(p)],
\end{equation}
where $B_i(p)$ and $L_i(p)$ are for the $i$th transmitter. We plan to explore general representations for multiple transmitters in 
future works.

\begin{figure}[t]
	\centering
	\subfigure[Mean Power (watt).]{
		\label{img1}
		\includegraphics[height=2.5cm,width=3.3cm]{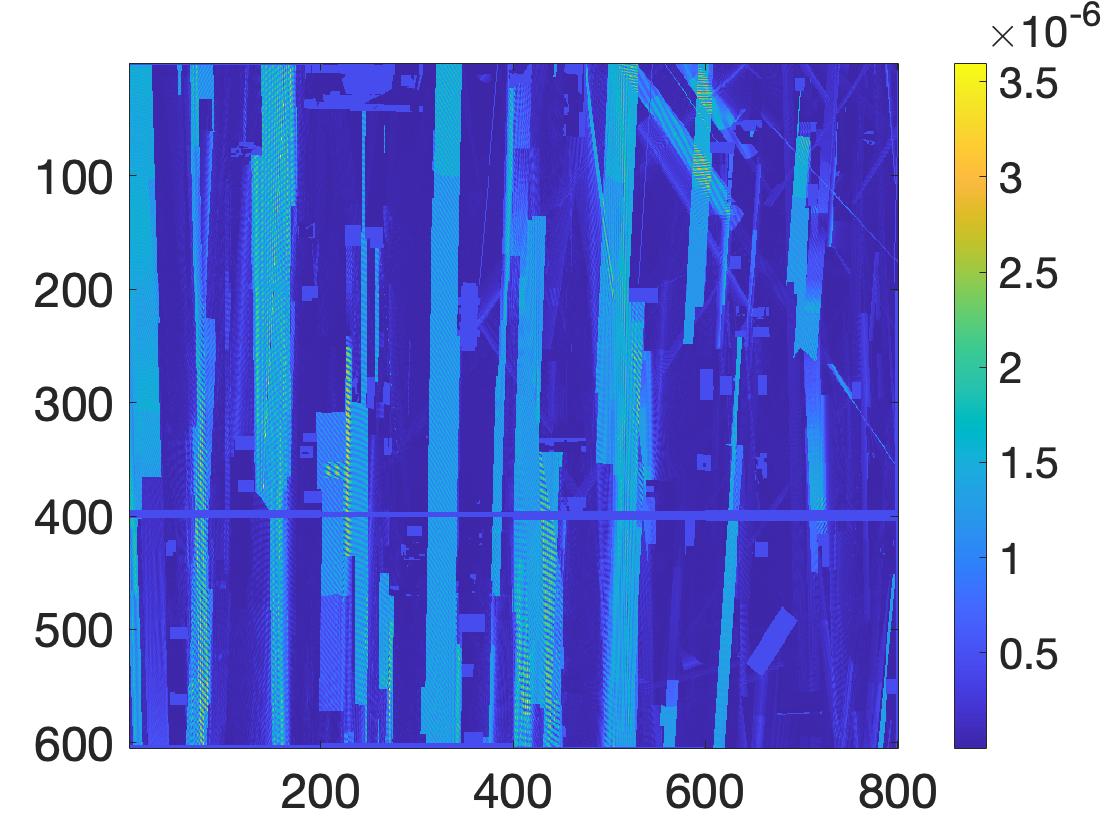}}
	\hspace{0.8cm}
	\subfigure[Satellite]{
		\label{img2}
		\includegraphics[height=2.5cm,width=2.5cm]{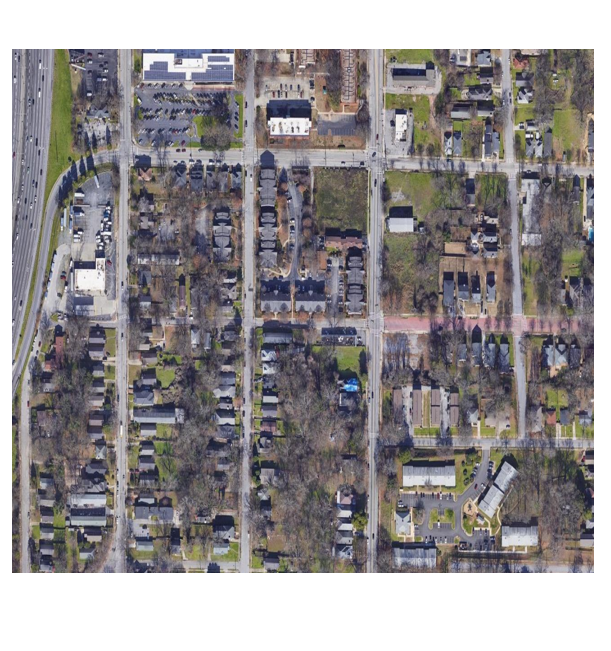}}
	\vspace{-2mm}
	\caption{Illustration of APL dataset.}
	\vspace{-3mm}
	\label{data_img}
\end{figure}
\begin{figure}[t]
	\centering
	\subfigure[]{
		\label{img11}
		\includegraphics[height=2cm,width=2.7cm]{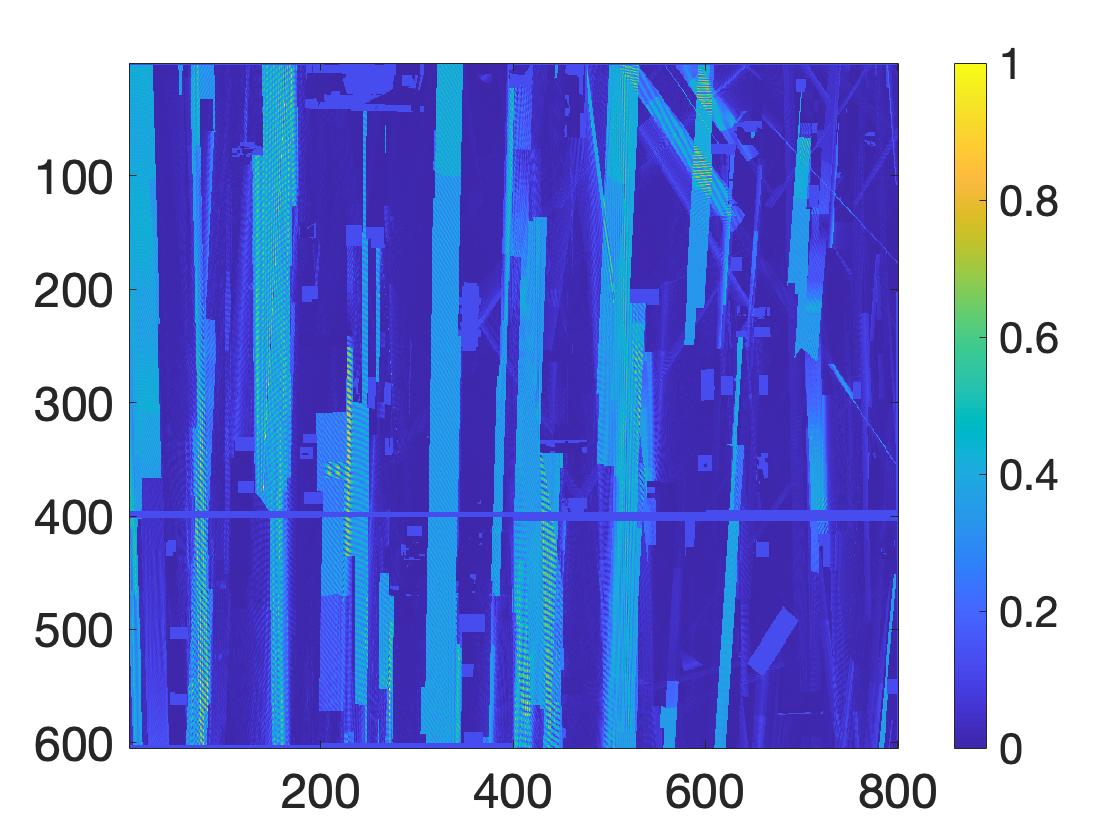}}
	\subfigure[]{
		\label{img21}
		\includegraphics[height=2cm,width=2.7cm]{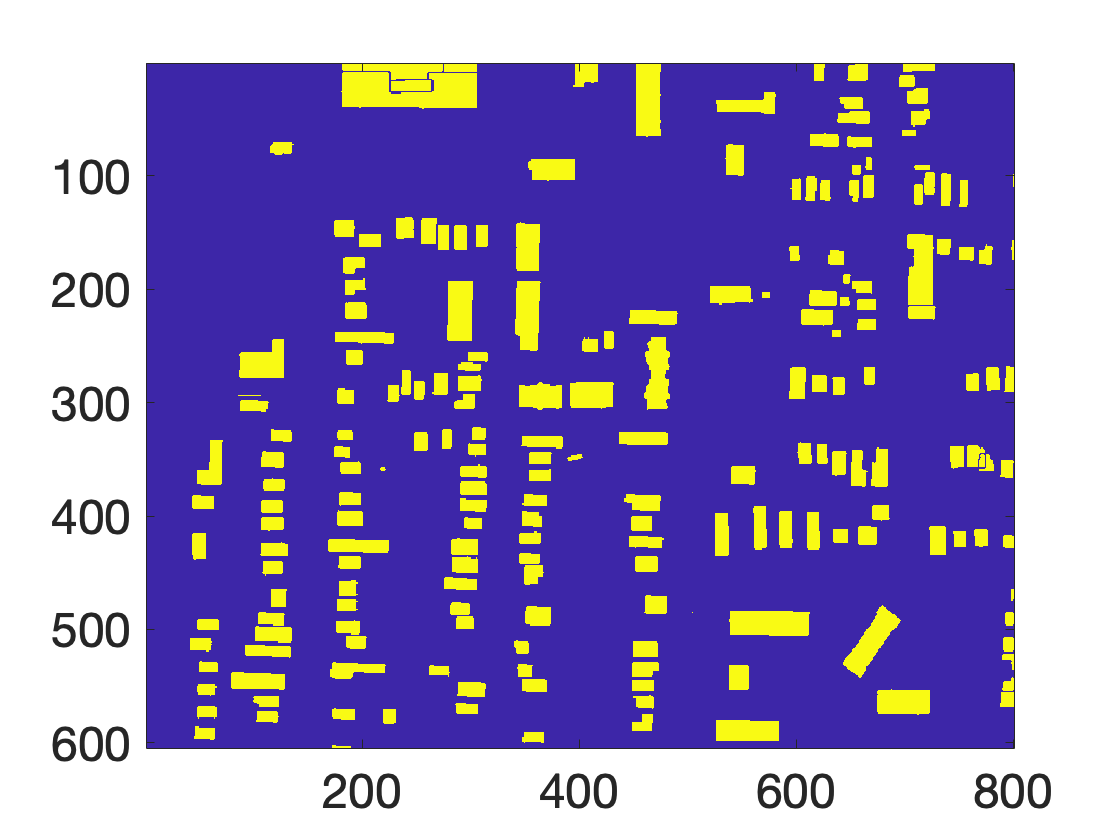}}
	\subfigure[]{
		\label{img31}
		\includegraphics[height=2cm,width=2.7cm]{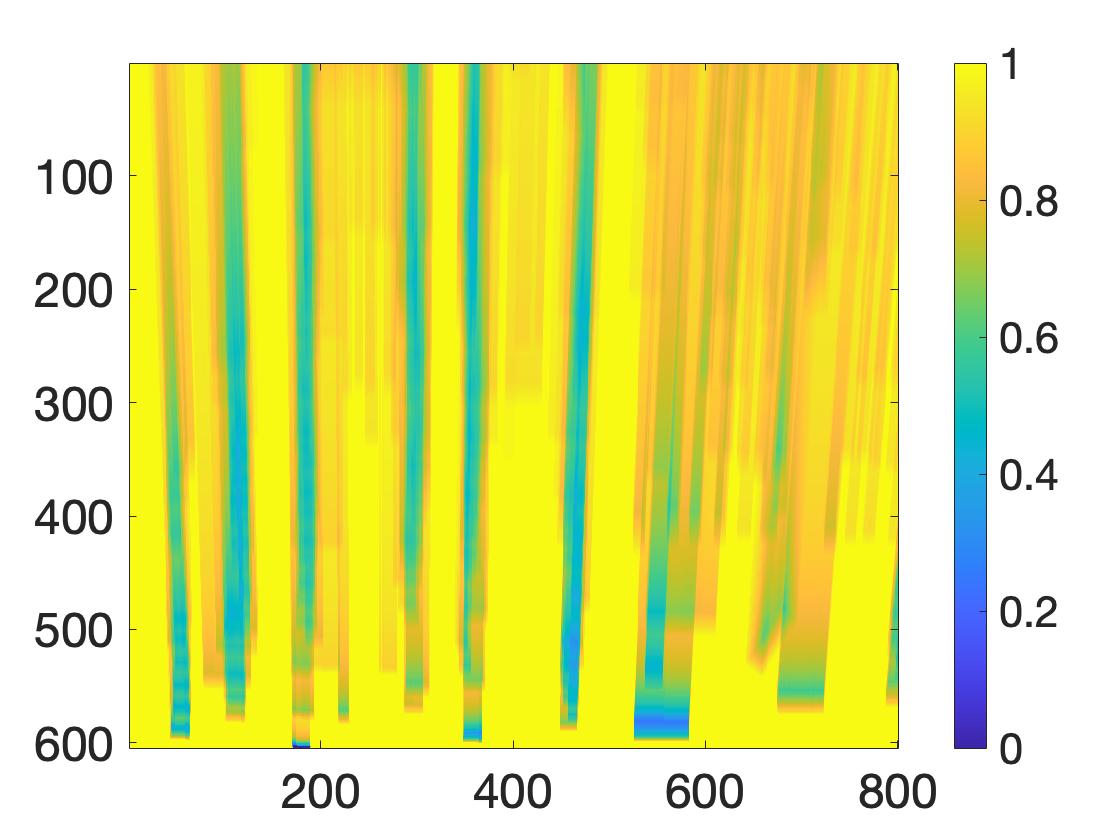}}
	\vspace{-2mm}
	\caption{Preprocessing of APL Dataset: a) Normalized radio map; b) Segmented buildings; and c) Block term in priority.}
	\label{data_img1}
	\vspace{-3mm}
\end{figure}

\subsubsection{Estimation of Missing Measurement}
After selecting the patch $\Psi_q$ with highest priority, the next step is to estimate the missing measurement values from identified regions. In this part, we introduce two exemplar-based approaches as follows:
\begin{itemize}
	\item Estimation based on exemplar copy (EPC):
	Copying values from similar patches in the observed region at the same indices is a widely-used approach to fill the missing regions \cite{c15}. Here, we also consider exemplar-based copying to reconstruct the radio map. Let $\Psi_q$ be the $n\times n$ patch selected by $P(p)$. We first find the most similar exemplar patch $\Psi_s$ from the observed region according to
	\begin{equation}
		\Psi_s=\arg \min_{\Psi_w, w\in \Phi}\sum_{i\in \Phi}[(\Psi_w)_i-(\Psi_q)_i]^2,
	\end{equation}
	where $(\Psi)_i$ is the PSD value at position $i$ within the patch $\Psi$. We then fill the missing value as
	\begin{equation}
		(\tilde \Psi_q)_i=\left \{
		\begin{aligned}
		(\Psi_q)_i	&\quad \quad i\in\Phi\\
			(\Psi_s)_i&\quad\quad i\in\Omega
		\end{aligned}
		\right..
	\end{equation}
	\item Estimation based on dictionary learning (EPD): Generating a dictionary from observations, one can optimize a sparse vector to combine the code-words in the dictionary to estimate missing values in the patches \cite{b16}. After selecting $n\times n$ patch $\Psi_q$, we can randomly pick $W$ patches from $\Phi$ and generate a dictionary $\mathbf{A}\in\mathbb{R}^{n^2 \times K}$ containing $K$ normalized code-words via K-SVD \cite{b17} or matching pursuit \cite{b18}. Reshaping patch $\Psi_q$ as a vector $\mathbf{x}_q$, we formulate dictionary learning as follows:
	\begin{equation}
		\tilde{\bm{\beta}}=\arg \min_{\bm{\beta}} ||(\mathbf{x}_q)_\Phi-\mathbf{A}_\Phi \bm{\beta}||^2_2+\lambda||\bm{\beta}||_1,
	\end{equation}
	where $\bm{\beta}\in\mathbb{R}^{K\times 1}$ is a sparse vector and $(\mathbf{x}_q)_\Phi$ is the observed part in $\Psi_q$. From the optimal $\bm{\beta}$, we reconstruct the radio map as 
	\begin{equation}
	(\tilde \Psi_q)_i=\left \{
	\begin{aligned}
	(\Psi_q)_i	&\quad \quad i\in\Phi\\
	(\mathbf{A}\bm{\beta})_i&\quad\quad i\in\Omega
	\end{aligned}
	\right..
	\end{equation}
\end{itemize}
In general, exemplar-based copying performs better when the radio map has regular, continuous patterns, while the dictionary learning performs better when the environment is more complex. See more discussions in Section \ref{exp}. Other potential ways to estimate the missing values include subspace learning \cite{b19} and graph learning \cite{b20}.

\section{Experiment Results} \label{exp}
In this section, we present test results to demonstrate the efficacy of the proposed methods.

\subsection{Data Information and Preprocessing}
Our test is based on the APL dataset which was generated from Wireless inSite Software \cite{a1} with Light Detection and Ranging (LIDAR) information of a select region in Atlanta, Georgia, USA. The LIDAR data used for the simulation has a 1-meter resolution.
The APL dataset contains a transmitter (Tx) and distributed single-antenna receivers in a 10-block area. The TX antenna is a uniform square array of $16\times 16$ elements, spaced at 0.5 wavelength. The TX is located at latitude/longitude of 33.689/-84.390. The antenna height is 201 meters, and the frequency used is 2660 MHz. The receiver antennas assumed
a height of 2.01 meters and uniformly
spaced by 0.8 meters. The location of the observed area is at 33.7283$\sim$33.7327 in latitude and -84.3923$\sim$-84.3854 in longitude. To generate the radio map from APL data, we average antenna gains from TX for each data point and conform it to a $604\times 800$ grid, i.e., $U(\mathbf{Z})\in \mathbb{R}^{604 \times 800}$, where the grid resolution (each $1\times 1$ block) is in 0.8 meters. Note that some original points might be arranged to shared positions in the grid during this process. For those data, the values are further averaged in the shared locations. 
The mean power in $\mathbf{Z}$, together with its satellite image, are presented in Fig. \ref{data_img}.
For convenience, we linearly normalize the radio map between $0\sim 1$. Note that the original radio map can be transformed without loss from the normalized one, and their pattern are exactly the same shown as Fig. \ref{img11}. Based on the satellite map, we segment the buildings against the background and calculate the block term in the priority by Eq. (\ref{btm}), shown as Fig. \ref{img21} and Fig. \ref{img31}, respectively.

\begin{figure}[t]
	\centering
	\subfigure[Scenario 1]{
		\label{ss1}
		\includegraphics[height=2.5cm,width=3.3cm]{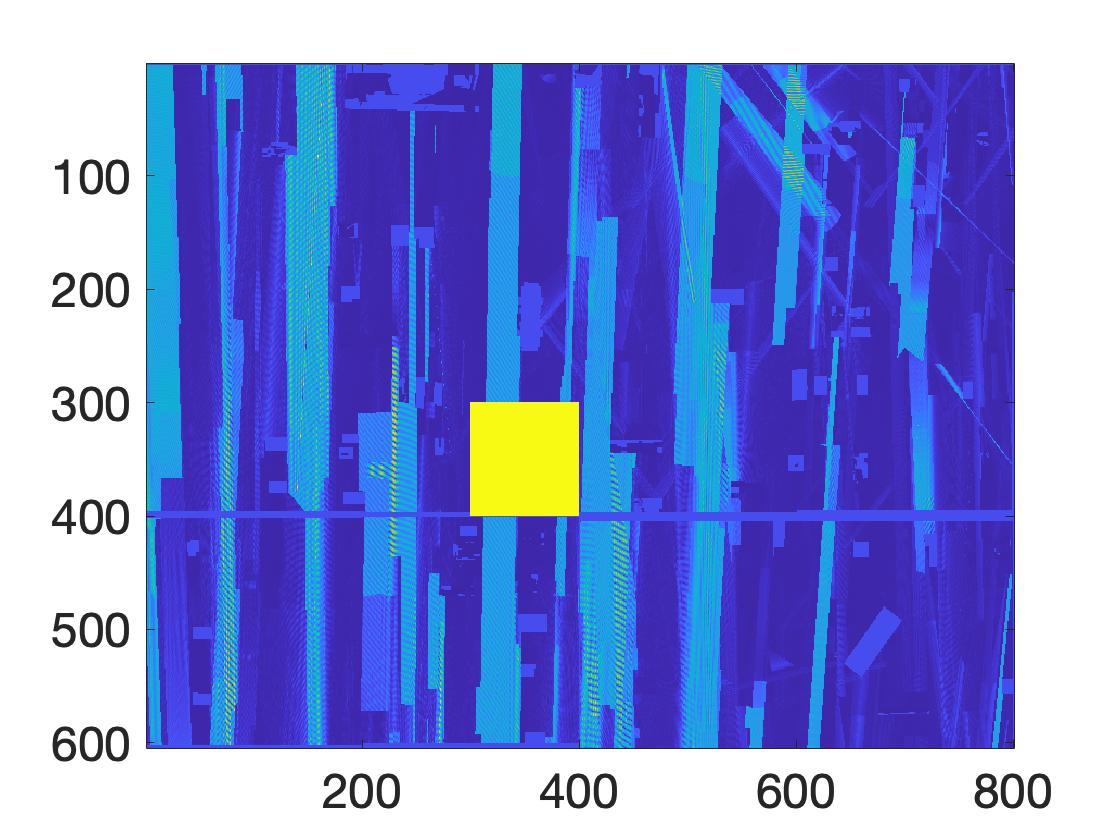}}
	\hspace{0.8cm}
	\subfigure[Scenario 2]{
		\label{ss2}
		\includegraphics[height=2.5cm,width=3.3cm]{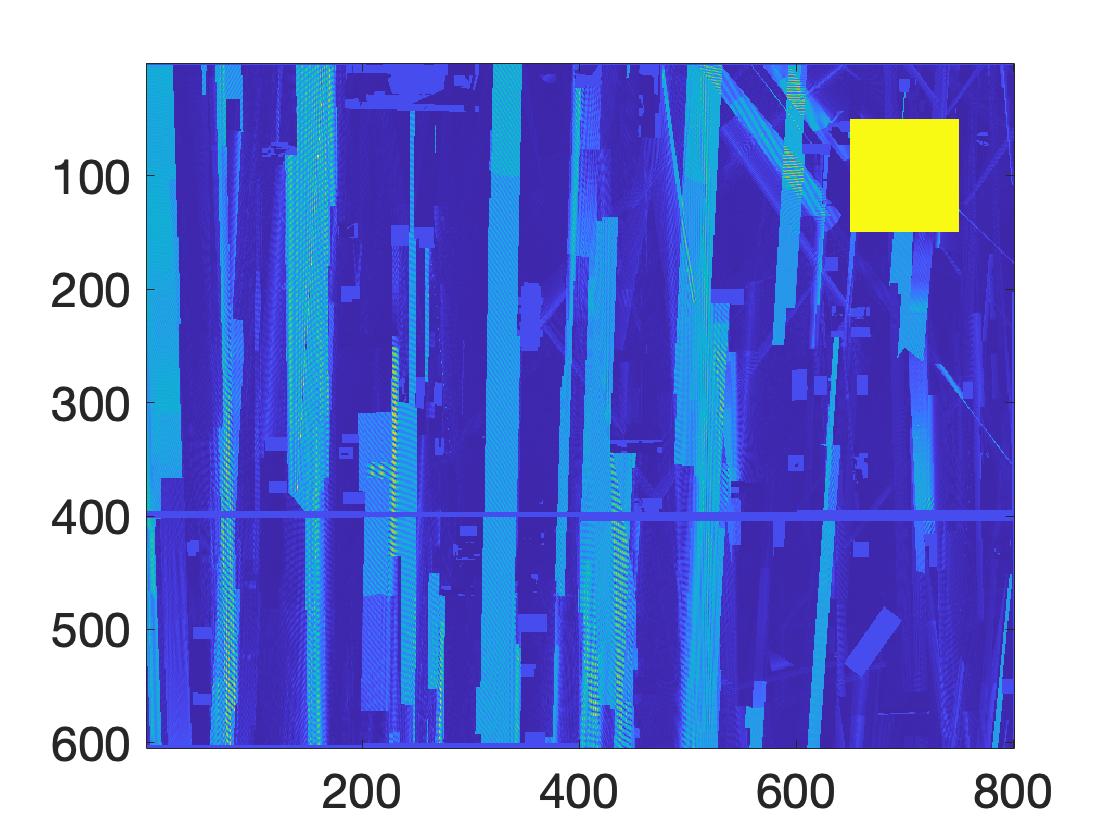}}
	\vspace{-2mm}
	\caption{Selected Areas to Test Performance: a) Scenario with regular neighborhood pattern; and b) Scenario with complex neighborhood pattern. The restricted/inaccessible areas $\mathbf{Z}_p$ with size $100\times 100$ are marked in yellow.}
	\vspace{-5mm}
	\label{sss}
\end{figure}

\subsection{Performance in Selected Areas}\label{psa}
To measure performance, we first consider two specific scenarios, i.e., one with regular neighborhood pattern and one with complex neighborhood pattern shown in Fig. \ref{sss}. In both scenarios, we considered a restricted/inaccessible areas $\mathbf{Z}_p$ with area size $100\times 100$ in grid. The PSD in the whole restricted areas (marked as yellow in Fig \ref{sss}) is unavailable, which we reconstruct from other observed parts in $\mathbf{Z}$.

We compare our methods with Model-based Interpolation (MBI) \cite{b5}, Radial Basis Function (RBF) Interpolation \cite{b8}, Label Propagation (LP) \cite{b10}, Exemplar-based Inpainting (EI) \cite{c15}, and Dictionary Learning (DL) \cite{b16}.
MBI and RBF are interpolation methods based on distances. EI and DL are image inpainting approaches without using radio propagation knowledge. For LP, we incorporate satellite images and information
on distance to transmitter as features.
For our proposed method, we consider three setups: 1) texture priority together with block term under exemplar-based copy (EBC); 2) complete priority with all 4 terms under exemplar-based copy (EPC); and 3) complete priority with all 4 terms under exemplar-based dictionary learning (EPD). For image inpainting methods and our proposed methods, we select patch size of $\Psi_p$ as $21\times 21$ for fair comparison. For methods related to dictionary learning, we set the number of code-words to $K=500$. 
We apply K-SVD \cite{b17} to generate the dictionary.

\begin{figure*}[t]
	\centering
	\subfigure[Reconstructed Results for Scenario 1.]{
		\label{r1}
		\includegraphics[width=7in]{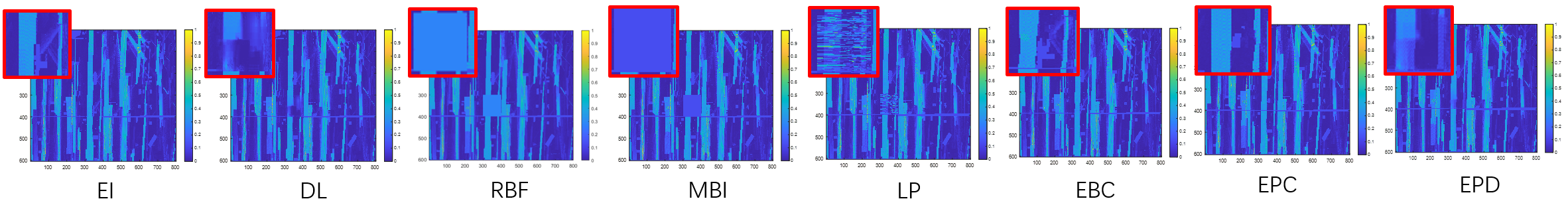}}\\
	\vspace{-2mm}
	\subfigure[Reconstructed Results for Scenario 2.]{
		\label{r2}
		\includegraphics[width=7in]{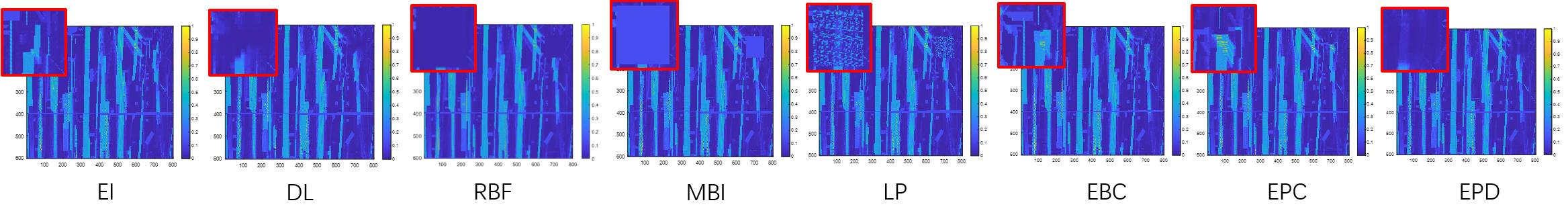}}
	\vspace{-3mm}
	\caption{Visualized Results in Selected Areas: (a) and (b) describe the regular and complex area, respectively; the results in red blocks are zoom-in presentations.}
	\label{rr}
	\vspace{-3mm}
\end{figure*}

\begin{table*}[t]
	\caption{Numerical Results in Selected Areas.}
	\centering
	\begin{tabular}{|l|l|l|l|l|l|l|l|l|}
		\hline
		& EI     & DL     & RBF    & MBI    & LP     & EBC    & EPC             & EPD             \\ \hline
		MSE in Scenario 1 & 0.0092 & 0.0152 & 0.0448 & 0.0271 & 0.0327 & 0.0088 & \textbf{0.0038} & 0.0096          \\ \hline
		MSE in Scenario 2 & 0.0258 & 0.0158 & 0.0217 & 0.0173 & 0.0306 & 0.0227 & 0.0152          & \textbf{0.0136} \\ \hline
	\end{tabular}
\vspace{-4mm}
\label{tt}
\end{table*}

The visualization results are shown in Fig \ref{rr}, and the corresponding numerical results are shown as Table \ref{tt}. Here, we define
$\mbox{MSE}=\frac{1}{m}\sum_{i=1}^m (x_i-\tilde{x}_i)^2$,
 where $\tilde {x}_i, i=1,\cdots,m$ are the estimated radio map.
 Shown as Fig. \ref{rr}, model-based MBI fails to estimate the radio map in the restricted/inaccessible area since the power spectrum in this dataset is over smaller
 distance variation from the transmitter but is more sensitive
 to the surrounding environment as 
 seen from Fig. \ref{data_img}. The RBF interpolation also fails to reconstruct missing segments and fills missing radio map with 
 similar values since the observations are uneven, especially near the center of the restricted/inaccessible areas. 
 For learning-based LP, the results display strong noises since the training samples from satellite images are noisy. Compared to the image inpainting methods, the proposed methods based on radio propagation priority show superior performance, since propagation information can enhance the features and textures. As shown in Fig. \ref{img31},  propagation priority terms favor the vertical direction to fill the region, which match the distribution of spectrum pattern in Fig. \ref{img1}. In our proposed methods, copy-based estimations display sharper features while dictionary learning based estimation provides more robust but blurred results. In the first scenario with regular nearby patterns close to the main road, EPC displays significant improvement since the vertical patterns therein is clear and similar. In the second scenario near buildings and trees, EPC sometimes over-estimates some regions from neighborhoods while EPD displays more robustness. The numerical results in Table \ref{tt} are consistent with the visualization results. Thus, one can determine whether EPC or EPD should be selected for estimation depending on the variations of
 the nearby environment.

\subsection{Overall Performance for Different Area Sizes}
\begin{figure*}[t]
	\centering
	\subfigure[MSE.]{
		\label{mse}
		\includegraphics[height=3.1cm]{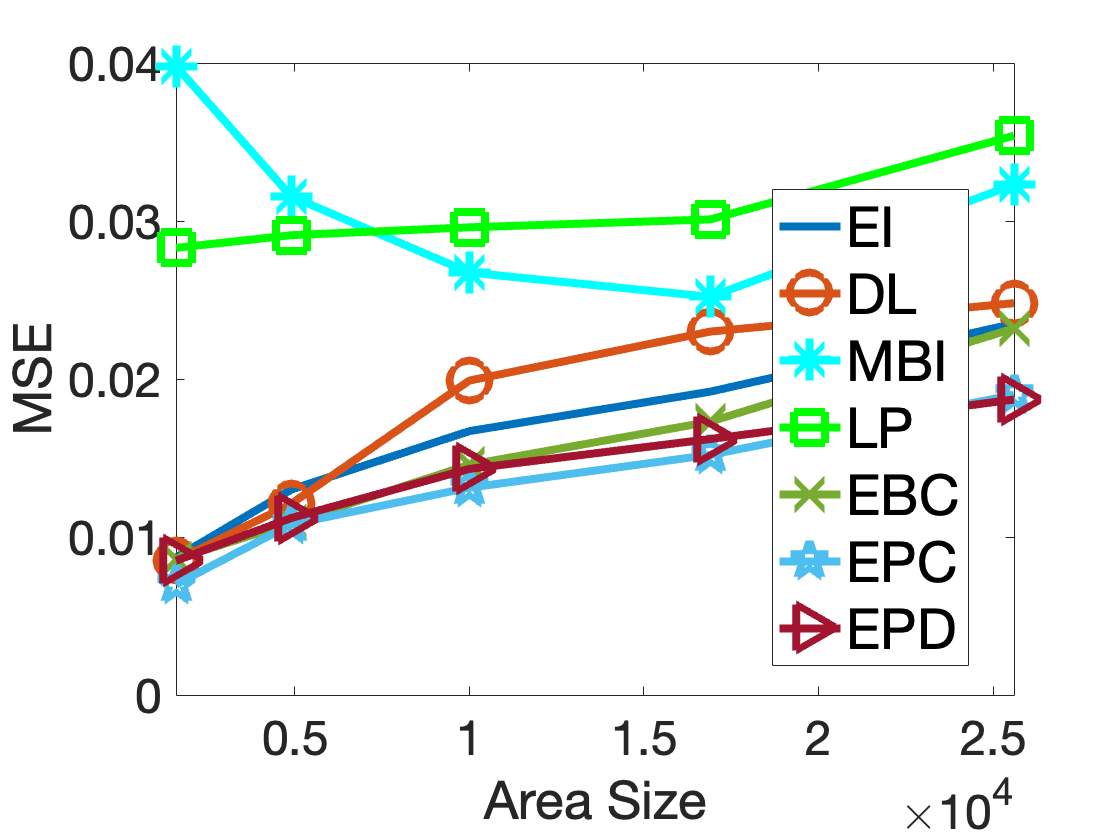}}
	\hspace{5cm}
	\subfigure[NE]{
		\label{ne}
		\includegraphics[height=3.1cm]{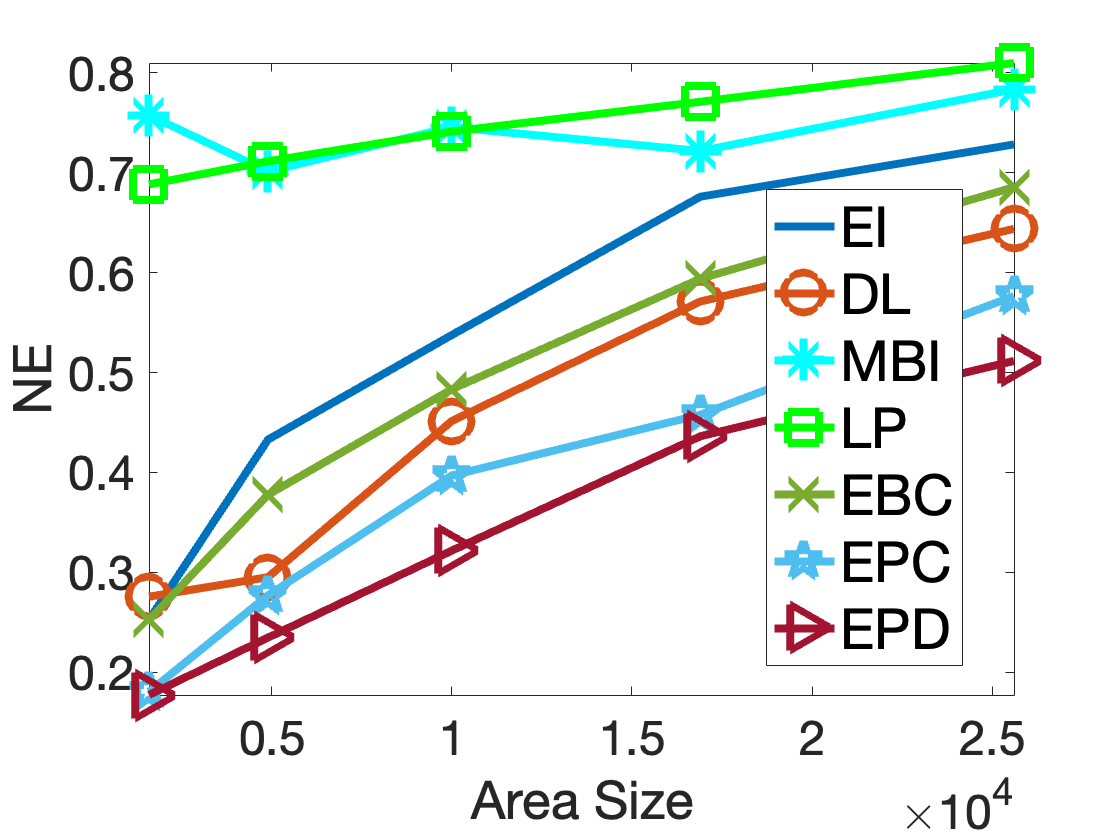}}
	\caption{Numerical Results in Different Area Sizes}
	\label{err}
	\vspace{-5mm}
\end{figure*}

We further examine the overall radio map estimation performance for different area sizes. In this test, we compare different
methods 
for restricted/inaccessible areas of various area sizes, i.e., $30\times 30$, $70\times 70$, $100\times 100$, $130\times 130$, and $160\times 160$. For each size, we randomly generate 10 restricted/inaccessible areas as the target region within Fig. \ref{img1}. We then calculate the mean error of different generated areas to implement the comparison.
In addition to MSE, we define a normalized error ({NE}), i.e.,
$\mbox{NE}=\frac{\sum_{i=1}^m(x_i-\tilde{x}_i)^2}{\sum_{i=1}^m x_i^2}$.
The results are shown in Fig. \ref{err}. Since MBI fails to capture the spectrum patterns in small-scale areas, it displays steadily poor result. For other methods, radio map error increases as the area size grows. This is intuitive since  neighborhood information and observations become
more limited and uneven for larger restricted/inaccessible areas, especially near the center of the restricted/inaccessible area. Our proposed methods are better than traditional inpainting and LP approaches, demonstrating the important impact of the proposed radio propagation priority. EPC and EPD show similar MSE results while EPD generate better 
NE than EPC. 
The results indicate
that EPC works better in some special scenarios whereas EPD is more robust regardless
of the power in the restricted/inaccessible areas. The conclusions are similar to Section \ref{psa} and further demonstrate the benefits of the proposed method.

\subsection{Guidelines of Parameter Selection}

In this part, we consider the proposed methods under different parameters to develop selection guidelines. We first evaluate the impact of patch sizes in $\Psi_p$ for EPC and EPD in Table. \ref{psz}. 
For EPC, we test a randomly selected $100 \times 100$ restricted/inaccessible area. 
For EPD, we set $K=1000$ for the dictionary and test a
$40 \times 40$ restricted/inaccessible area.
Patch size selection
is a trade-off between the global information and local observations. For a larger patch size, uncertainty 
grows with more global information considered.
From the results, we determine
a suitable patch size
around 15$\sim$21. We also test EPD with different dictionary sizes in Table \ref{ddz}, which shows that a larger $K$ can
achieve better performance.

\begin{table}[t]
	\centering
	\caption{MSE in Different Patch Size}
	\begin{tabular}{|l|l|l|l|l|l|}
		\hline
		Patch Size                                                                                               & 9                       & 15                      & 21                      & 27                      & 33                      \\ \hline
		MSE for EPC       & {0.0177} & {0.0132} & {0.0152} & {0.0163} & {0.0205}                  \\ \hline
		MSE for EPD   & {0.0020} & {0.0018} & {0.0025} & {0.0027} & {0.0034} \\
		 \hline
	\end{tabular}
\label{psz}
	\vspace{-3mm}
\end{table}

\begin{table}[t]
	\centering
	\caption{MSE for EPD with Different Dictionary Size}
	\begin{tabular}{|l|l|l|l|l|}
		\hline
		K (Patch size=15) & 500    & 1000   & 1500   & 2000   \\ \hline
		MSE     & 0.0026 & 0.0025 & 0.0020 & 0.0020 \\ \hline
	\end{tabular}
\label{ddz}
\vspace{-3mm}
\end{table}

\section{Conclusion}
In this work, we introduce an exemplar-based approach to
wireless radio map reconstruction in the cases of missing measurement. More specifically, we proposed a propagation-based priority to determine the filling direction based on PSD pattern and radio properties. We then
introduced two new schemes for patch estimation. The experimental results demonstrate the efficiency of the propagation-based priority to capture the PSD patterns and the power of our proposed methods in radio map reconstruction for missing areas, which make further spectrum access and management more reliable for such restricted/inaccessible areas.

%\section*{Acknowledgment}

\end{document}